\documentclass{desyproc}
\begin{document}
\title{Heavy quarkonium production at the LHC in the framework of NRQCD and parton Reggeization approach}

\author{{\slshape Maxim Nefedov$^1$, Vladimir Saleev$^1$}\\[1ex]
$^1$ Samara State University, Academic Pavlov st. 1, 443111 Samara,
Russia}

\contribID{xy}

\desyproc{DESY-PROC-2013-03}
\acronym{HQ2013} 

\maketitle

\begin{abstract}
  Heavy quarkonium production in the framework of the nonrelativistic quantum chromodynamics and
  leading order of the parton Reggeization approach at the Tevatron and LHC Colliders is discussed.
  The new results, which are reviewed in this report, include the comparison with recent data on $\chi_{c1,2}$
  production from ATLAS Collaboration and special discussion of heavy quarkonium polarization issues in the considered framework.
\end{abstract}

\section*{Introduction and basic formalism.}

  Production of heavy quarkonia at hadron colliders is the unique laboratory for the studies of the interplay between
  theory of perturbative hard subprocess and models of nonperturbative hadronization in the hadronic collision.
  The hope to understand the hadronization stage is associated with the nonrelativistic nature of the problem, which
  allows one to organize theoretical predictions in a form of double expansion in powers of strong coupling constant
   $\alpha_s$ and relative heavy quark velocity $v$.

 In the  nonrelativistic Quantum Chromodynamics (NRQCD)  one can  factorize the effects of short and long distances
 in the cross section of heavy quarkonium production as follows~\cite{NRQCD, NRQCDMaltoni}:
  \begin{equation}
  d\hat \sigma (I \to {\cal H})=\sum\limits_{n} d\hat \sigma (I \to Q\bar{Q}[n]){\langle {\cal O}^{\cal
H}[n]\rangle},
  \end{equation}
   where the sum is over the possible intermediate states of heavy quark-antiquark ($Q\bar{Q}$) pair
    $[n]=^{2S+1}L_J^{(1,8)}$, with definite spin $S$, orbital momentum $L$, total angular momentum $J$ and color-singlet (CS) $^{(1)}$ or
    color-octet (CO) $^{(8)}$ quantum numbers. Factor $d\hat \sigma$ is the partonic cross section of production of the state $Q\bar{Q}[n]$
    from the partonic initial state $I$, and $\langle {\cal O}^{\cal H}[n]\rangle$ are the nonperturbative matrix elements (NMEs),
     describing the transition of the intermediate state $Q\bar{Q}[n]$ to the heavy quarkonia ${\cal H}$.
     In our calculations, the normalization of NMEs and cross section is chosen the same as in the Ref.~\cite{NRQCDMaltoni}.

     According to the NRQCD velocity scaling rules~\cite{NRQCDvSR}, the following CS
     NMEs give the leading contribution to the production of quarkonia with the same spin-orbital quantum numbers:
     $
   \left\langle {\cal O^H}\left[^3S_1^{(1)}\right]\right\rangle ,\ \left\langle{\cal  O^H}\left[^3P_J^{(1)}\right]\right\rangle .
     $
     The CO NMEs -- $
     \left\langle {\cal O^H}\left[^1S_0^{(8)}\right]\right\rangle,  \ \left\langle {\cal O^H}\left[^3S_1^{(8)}\right]\right\rangle ,
     \ \left\langle{\cal  O^H}\left[^3P_J^{(8)}\right]\right\rangle$,
      give the next to leading contributions in $v$. CS NMEs could be expressed through the quarkonia wave function,
      and then calculated in the potential quark model. CO NMEs describe the transition of CO $Q\bar{Q}$ pair into
      quarkonia by radiation of the soft gluons, and hence could not be computed neither in perturbative QCD, nor in the potential quark models.
       The only option, available so far, is to fit this NMEs to reproduce experimental data.

      The latter means, that the hard part of the cross-section should be calculated as precisely as possible,
      to get physically meaningful results. Nowadays the complete next-to-leading-order (NLO) results for inclusive heavy quarkonia production
      are available~\cite{KniehlPSI, GWZUpsi}. However, fixed order calculations are applicable only in the region of $p_T\gg 2m_{Q}$.
      In the region of small $p_T$, the resummation of the large logarithms $log(m_Q/p_T)$ is needed to obtain reliable predictions.
      The existing calculations, based on the small-$p_T$ resummation procedure, see e. g.~\cite{smallPTR}, are restricted
      to the region $p_T\ll 2m_Q$, and require matching with the fixed order calculations at higher $p_T$. So, the approach,
      which takes into account both small and high $p_T$ regions on the same grounds is needed to obtain the values of CO NMEs.

      Such approach could be designed, using the $k_T-$factorization~\cite{TMDf}, which naturally regularizes the small-$p_T$ divergences,
      present in the fixed-order calculations in the collinear PM.

      The dominating contribution to the inclusive heavy quarkonium production at hadron colliders, comes from the gluon fusion subprocess.
       The cross section for this process in the framework of $k_T-$factorization is represented as a convolution of
       unintegrated parton (gluon) distribution functions (PDFs) in a proton $\Phi_g^p\left(x,t,\mu^2\right)$ with the partonic cross section.
        Unintegrated PDF depends on the longitudinal momentum fraction $x$, the virtuality of the parton $t=-q_T^2={\bf q}_T^2$,
        and the factorization scale $\mu$.

      Virtuality of the partons in the initial state of the hard subprocess, usually breaks the gauge invariance of the amplitude.
      However, it was shown~\cite{QMRK}, that in QCD at high energies, the so-called quasi-multi Regge kinematics dominates,
      when produced particles are arranged in clusters, strongly separated in rapidity. In this high-energy (Regge) limit,
      the gauge invariance condition holds for each of this clusters independently from the others, so the fields, carrying four-momentum
      between this clusters, are new gauge invariant degrees of freedom, accompanying the ordinary gluons and quarks in the effective
      field theory for the Regge limit of QCD \cite{LipatovEFT}. They are Reggeized gluons and Reggeized quarks~\cite{LipatovEFT,LipVyaz}.

      In our calculations, we rely on the assumption, that particles produced in the hard subprocess are well separated in rapidity from ones,
      produced at the evolution stage. Therefore, partons incoming to the hard subprocess are Reggeized, and we use the Feynman rules
      of Ref.~\cite{LipVyaz, FRgluons} to compute the hard scattering matrix elements. Matrix elements for the relevant
      $2\to 1$ and $2\to 2$ subprocesses have been obtained in Refs.~\cite{KSVcharm, KSVbottom, SVYadFiz}.

      Although, unintegrated PDFs are not so constrained as usual collinear PDFs, there exists the method to obtain unintegrated PDFs
      from the collinear ones, which showed stable and consistent results in many phenomenological applications, it is the
      Kimber-Martin-Ryskin (KMR) method~\cite{KMR}. Together with the parton Reggeization approach (PRA), this method was recently applied
       to describe dijet~\cite{dijets} and bottom-flavored jet~\cite{bjets} production, Drell-Yan lepton pair production~\cite{DY},
        single jet and prompt photon production~\cite{PPSJ} at the Tevatron and LHC.

\section{Charmonium and Bottomonium production.}
  Now we start the discussion of recent results in the phenomenology of heavy quarkonium production, obtained in the leading order of the NRQCD and PRA.
  In the Ref.~\cite{SNScharm}, it was shown, that it is possible to describe the latest LHC experimental data on the prompt charmonium
  production at the $\sqrt{S}=7$ TeV in a wide kinematical range ($2<p_T<20$ GeV and $|y|<3.5$) with a good accuracy, using the CO NMEs
  extracted from Tevatron data at the
$\sqrt{S}=1.8$ TeV and $1.96$ TeV \cite{KSVcharm,SNScharm}. The
fitted CO NMEs are also shown to be compatible with NLO collinear
parton model (PM) results of Ref.~\cite{KniehlPSI}.

\begin{figure}[hb]
\centerline{\includegraphics[width=0.4\textwidth]{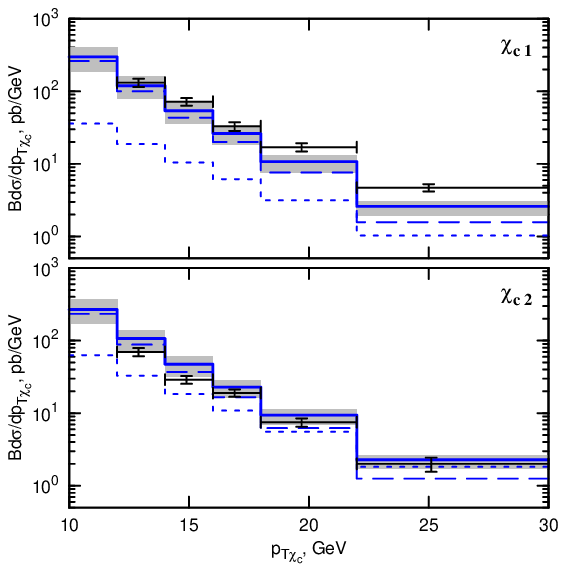}\includegraphics[width=0.4\textwidth]{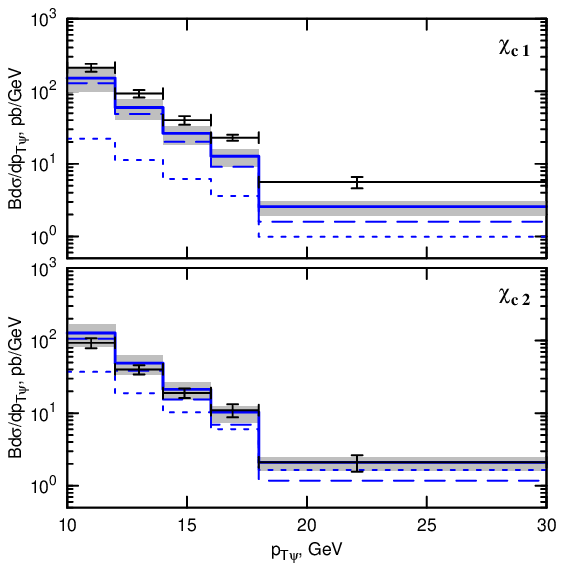}}
\caption{Transverse momentum spectra of prompt $\chi_{c1,2}$
production at $\sqrt{S}=7$ TeV, measured through their radiative
decay to $J/\psi$ mesons. Experimental data by ATLAS
collaboration~\cite{ATLAScn}. In the left panel, the reconstructed
spectrum over $p_{T\chi_c}$. In the right panel, the spectrum over
$p_{T J/\psi}$. Product of branching fractions $B=B(\chi_{c1,2}\to
J/\psi \gamma)B(J/\psi \to\mu^+\mu^-)$ is included. Dashed line is
the $^3P_J^{(1)}$ contribution, dotted line is the $^3S_1^{(8)}$
contribution, solid line is their sum.}\label{fig1}
\end{figure}

  Very recently, ATLAS collaboration has presented the measurement of the prompt and non-prompt $\chi_{c1}$ and $\chi_{c2}$ production
  in $pp$-collisions at $\sqrt{S}=7$ TeV~\cite{ATLAScn}. Comparison of the leading order (LO) PRA predictions
  with this new data is presented in the Fig.~\ref{fig1}.

  To produce the predictions in the Fig.~\ref{fig1}, we used the CO NMEs of Ref.~\cite{SNScharm},
  which where fitted to the Tevatron data at $\sqrt{S}=1.8$ TeV. In our calculations, $m_c=M_{J/\psi}/2$,
  and momentum squared of produced $Q\bar{Q}$ pair is equal to $M_{J/\psi}^2$ for all
  states. The proper treatment of the mass difference between the $Q\bar{Q}$
   pair and produced quarkonium, requires taking into account higher order relativistic corrections in $v$.

 To estimate $d\sigma/dp_{T\chi_c}$,
 we rescaled the transverse momentum by the mass ratio $M_{J/\psi}/M_{\chi_{cJ}}$, which corresponds to
  the approximation of collinear radiation of the decay photon in the limit $M_{\chi_{cJ}}-M_{J/\psi}\ll
  M_{J/\psi}\ll p_{TJ/\psi}$, as it was used in Ref.~\cite{GWZUpsi}.

  The $\Upsilon(nS)$ production in the LO PRA and NRQCD was studied in the first time in Ref.~\cite{KSVbottom},
  the detailed discussion of the recent LHC data is presented in the Ref.~\cite{SNSbottom}. It is pointed out,
  that the inclusion of the region of small $p_T$, greatly constrains the fit, and suppresses possible negative values of CO NMEs.
  Also, negative values of NMEs could not be advocated in our formally LO calculation.

\section{Heavy quarkonium polarization puzzle.}

  The study of the polarization of $S-$wave heavy quarkonia is very important for testing of the NRQCD factorization,
  since the soft gluon exchange at the hadronisation stage is belived to be not able to sufficiently change
  the polarization of the $Q\bar{Q}$ pair, produced in the hard scattering.

  The polarisation variables are defined through the angular distribution of the products of the decay
  ${\cal H}\to \mu^+\mu^-$ in the rest frame of heavy quarkonium ${\cal H}$:
  \begin{equation}
  \frac{d\sigma}{d\Omega}\sim 1+\lambda_\theta \cos^2(\theta)+\lambda_\varphi \sin^2(\theta) \cos(2\varphi)
   + \lambda_{\theta\varphi} \sin(2\theta) \cos(\varphi)\ ,
  \end{equation}
  where $\theta$ and $\varphi$ are polar and azimuthal angles of lepton ($\mu^+$) momentum in the some coordinate system, chosen in the rest frame of ${\cal H}$, and $\lambda_\theta$, $\lambda_\varphi$, $\lambda_{\theta\varphi}$ are polarization parameters.
   The issue of the choice of the coordinate system is important and widely discussed in the literature, see e. g.~\cite{POLfrms},
   here we use only the $s$-channel helicity frame.

\begin{figure}[hb]
\centerline{\includegraphics[width=0.35\textwidth]{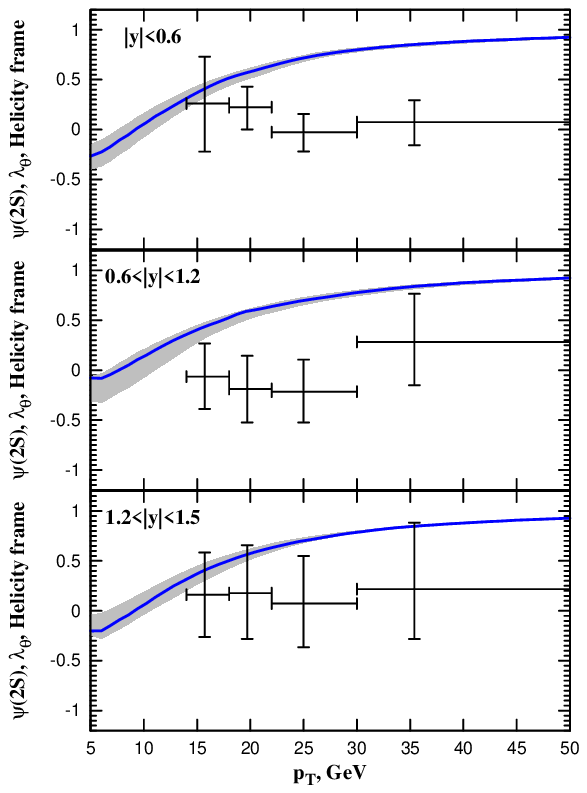}\
\ \ \ \
\includegraphics[width=0.35\textwidth]{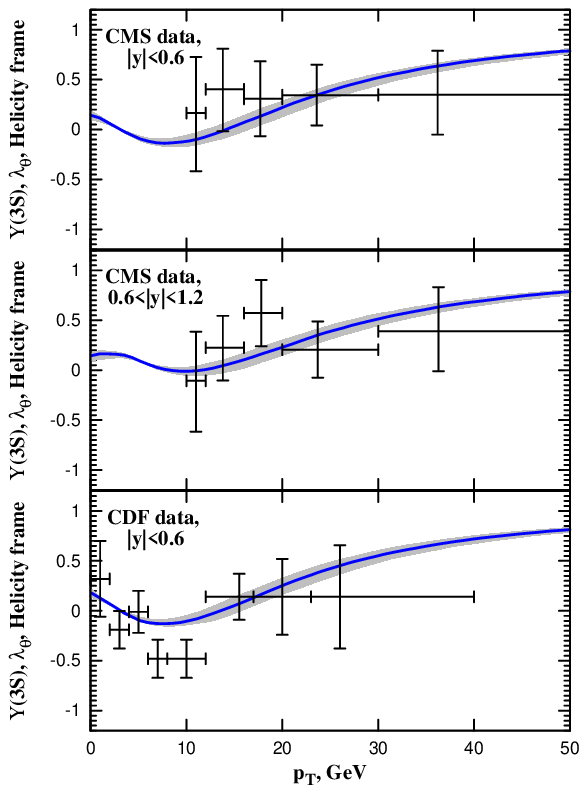}}
\caption{Left panel -- the polarization parameter
$\lambda_\theta$ as function of $p_T$ for the $\psi(2S)$ production
at $\sqrt{S}=7$ TeV. Experimental data are from CMS
Collaboration~\cite{CMSpol}. Right panel -- the polarization
parameter for the $\Upsilon(3S)$ production at $\sqrt{S}=7$ TeV (CMS
data~\cite{CMSpol}) and $\sqrt{S}=1.8$ TeV (CDF data~\cite{CDFpol}).
}\label{fig2}
\end{figure}

  In the Fig.~\ref{fig2} we present the comparison of the LO PRA predictions on polarization parameter $\lambda_\theta$ for $\psi(2S)$
  and $\Upsilon(3S)$ states with the recent experimental data by CMS~\cite{CMSpol} and CDF~\cite{CDFpol} Collaborations.
  We choose these states, because in our model they are produced directly, and give the most clear test of the production mechanism.
   From Fig.~\ref{fig2} we can conclude, that our prediction for $\Upsilon(3S)$ polarization is in a good agreement with experimental data.
    The situation here is very similar to the results of Ref.~\cite{GWZUpsi}, obtained in the NLO of collinear PM.
    In contrast, the polarization of $\psi(2S)$ is not described.
    Because of the domination of the CO contributions at high $p_T\gg M_{J/\psi}$, theory predicts the strong transversal polarization of $\psi(2S)$,
    while experimental data are compatible with zero polarization. The same disagreement is observed in the NLO PM calculations~\cite{GWZUpsi}.
    Together with the observed inconsistency of the CO NMEs, obtained as a result of the global fit on cross section data,
    with the data on prompt $J/\psi$ polarization~\cite{KBpol}, this result leads to the famous charmonium polarization puzzle.
     Attempts to resolve this puzzle require careful study of feeddown contributions and higher-order processes,
     such as $p+p\to J/\psi+c+\bar{c}+X$, which can sufficiently contribute at high
     $p_T$. In case of bottomonium production, NRQCD agrees with
     experimental data as for $p_T-$spectra as for polarization
     parameters. It means that $b-$quark mass is sufficiently
     large to suppress relativistic corrections and
     nonperturbative effects during the hadronization.

\section{Acknowledgments}

We are grateful to the Organizing Committee for kind hospitality
during the HQ-2013 Workshop. The work of M.~N. was supported by the Grant of President of Russian Federation MK-4150.2014.2. The work of M.~N. and V.~S. was supported in part by the Russian Foundation for Basic Research through Grant 14-02-00021.


\begin{footnotesize}



%

\end{footnotesize}


\end{document}